\documentstyle[12pt,psfig]{article}

\newcommand{\AmS}{{\protect\the\textfont2
\renewcommand{\thesection}{\Roman{section}}
  A\kern-.1667em\lower.5ex\hbox{M}\kern-.125emS}}
 
\begin{document}
\rightline {DFTUZ 98/04}
\vskip 2. truecm
\centerline{\bf First Order Phase Transition in 
Finite Density QCD}
\centerline{\bf using the modulus of the Dirac Determinant.}
\vskip 2 truecm
\centerline { R. Aloisio$^{a,d}$, V.~Azcoiti$^b$, G. Di Carlo$^c$, 
A. Galante$^{a,c}$, A.F. Grillo$^d$}
\vskip 1 truecm
\centerline {\it $^a$ Dipartimento di Fisica dell'Universit\'a 
dell'Aquila, L'Aquila 67100 (Italy).}
\vskip 0.15 truecm
\centerline {\it $^b$ Departamento de F\'\i sica Te\'orica, Facultad 
de Ciencias, Universidad de Zaragoza,}
\centerline {\it 50009 Zaragoza (Spain).}
\vskip 0.15 truecm
\centerline {\it $^c$ Istituto Nazionale di Fisica Nucleare, 
Laboratori Nazionali di Frascati,}
\centerline {\it P.O.B. 13 - Frascati 00044 (Italy). }
\vskip 0.15 truecm
\centerline {\it $^d$ Istituto Nazionale di Fisica Nucleare, 
Laboratori Nazionali del Gran Sasso,}
\centerline {\it Assergi (L'Aquila) 67010 (Italy). }
\vskip 3 truecm

\centerline {ABSTRACT}
\vskip 0.5truecm

\noindent
We report results of  simulations of strong coupling, finite density 
QCD obtained within a MFA inspired approach where the fermion determinant 
in the integration measure is replaced by its absolute value. Contrary 
to the standard wisdom, we show that within this approach a clear signal for 
a first order phase transition appears with a critical chemical potential 
in extremely 
good agreement with the results obtained with the Glasgow algorithm. 
The modulus of the fermion 
determinant seems therefore to preserve some of
the relevant physical properties 
of the system. We also analyze the dependence of our results on the quark 
mass, including both the chiral and large mass limit, and the theory in 
the quenched approximation.
  
\vfill\eject

Non-perturbative investigations of QCD at finite temperature and density 
have received much attention in the last years. The aim of these 
investigations is to find the matter conditions in the early Universe 
and to get a clear insight into experimental signatures in the heavy-ion 
collision experiments. Even if considerable progress has been achieved 
in the investigations of QCD at finite temperature and zero chemical 
potential using the lattice approach, the present situation of the field at 
finite density is not so satisfactory. As is well known, the complex nature 
of the determinant of the Dirac operator at finite chemical potential, 
which makes it impossible to use standard simulation algorithms based on 
positive-definite probability distribution functions, has much delayed 
investigations on the full theory with dynamical fermions. On the other 
hand the quenched approximation, which has been extensively and successfully 
used in simulations of QCD at zero chemical potential, seems to have 
some pathological behaviour when applied to QCD at finite density 
\cite{BAR,KOG}.

We discuss in this paper some features connected to simulations in finite
density QCD. The main topic concerns the use of the absolute value of
the fermionic determinant. 
We show that, against some theoretical prejudices 
based on random matrix models \cite{MISHA}, the relevant physical features 
of finite density QCD seem to be preserved after taking the absolute value 
of the Dirac-Kogut-Susskind operator.

As well known, the partition function of finite density QCD 

\begin{equation}
{\cal Z} = 
\int [dU] e^{-\beta S_{G}(U)} \det \Delta(U,m,\mu).
\end{equation}

\noindent
can be written 
as the product of the following two contributions

\begin{equation}
{\cal Z} = {\langle\,e^{i\phi_{\Delta}}\rangle}_{||}
\int [dU] e^{-\beta S_{G}(U)} \left |\det \Delta(U,m,\mu)\right |.
\end{equation}

\noindent
where the first factor in (2) 

\begin{equation}
{{\langle\,e^{i\phi_{\Delta}}\rangle}_{||}} = 
\frac{\int [dU] e^{-\beta S_{G}(U)} 
e^{i\phi_{\Delta}}\left |\det \Delta(U,m,\mu)\right |} 
{\int [dU] e^{-\beta S_{G}(U)} \left |\det \Delta(U,m,\mu)\right |}
\end{equation}

\noindent
accounts for the mean value of the cosine of 
the phase of the fermion determinant computed with the probability 
distribution function of the pure gauge theory times the modulus of 
the fermion determinant. The second factor of (2) is just 
the partition function 
we will use along this work. The first factor in (2) gives a net contribution 
to the free energy density only in the case in which it falls off 
exponentially with the lattice volume. For random matrix models it has been 
shown that this is what happens \cite{MISHA} and this is the origin for the 
theoretical prejudices about the relevance of the phase in QCD. Early 
simulations of QCD with the absolute value of the fermion determinant 
in small lattices \cite{GOCK1} seem to corroborate these 
theoretical prejudices.

For the 
abelian model in 0+1 dimensions it can be shown also that the first factor 
of (2) plays a fundamental role \cite{GIBBS1} and this is also the case  
for four dimensional 
QED as follows from the fact that the abelian model shows no dependence on 
the chemical potential $\mu$ or in other words, from the absence of 
baryons in this model. Notwithstanding that, it can be shown that the 
phase of the determinant is completely irrelevant in 0+1 dimensional 
QCD \cite{LAT97} but unfortunately no analytical results on this subject 
are available 
for four dimensional QCD.

In order to check to what extent taking the 
absolute value of the fermion determinant in the integration measure is 
a good approximation for full QCD at finite baryon density, we  report 
here results for the number density as a function of 
the chemical potential $\mu$ 
at infinite gauge coupling, where 
more data are available in the literature \cite{BARKOG}, \cite{KARSCH}.
The results for larger $\beta$, in particular in the physically interesting
scaling region, will be presented elsewhere.
 
Our numerical simulations have been performed using a $MFA$ \cite{MFA} 
inspired approach.
The idea is to consider $\det \Delta$ or its absolute value 
as an observable. 
In this case  $\det \Delta$ is not in the integration 
measure, and one  avoids the 
problem of dealing with a complex quantity in the generation of 
configurations. This can be done in a (in principle) exact way 
by means of the 
 $MFA$ algorithm \cite{MFA} where the mean value of the determinant 
at fixed pure gauge energy is used to reconstruct an effective 
fermionic action as a function of the pure gauge energy only.
Up to 
now this method has been succesfully used in several models (at 
zero density),
where it allows free mobility in the $\beta-m_q$ plane, including 
the chiral limit \cite{MFA}.

We used the GCPF (Grand Canonical Partition Function) formalism
to write the 
fermionic determinant as a polynomial in the fugacity $z = e^\mu$ 
\cite{GIBBS2}:

\begin{eqnarray}
\det \Delta(U;m_q,\mu)&=&\det\left(G+e^\mu T+e^{-\mu} T^\dagger\right)=
z^{3V}\det\left(P(U;m_q)-z^{-1}\right) \nonumber \\
&=&\sum_{n=-3L_{s}^{3}}^{3L_{s}^{3}}  a_n z^{nL_t} 
\end{eqnarray}
where the propagator matrix is
\[ P(U;m_q) =  \left( \begin{array}{cc}
-GT & T \\
-T   &  0 \end{array} \right) \\
\]

in which $G$ contains the spatial links and the mass term, $T$ contains 
the forward temporal links \cite{GIBBS2} and $V$  is the lattice volume. 
Once fixed 
$m_q$, a complete diagonalization of the $P$ matrix allows to reconstruct 
$\det \Delta$ for any $\mu$. 
Due to the $Z(L_t)$ symmetry of the eigenvalues of  $P$ it is 
possible to write $P^{L_t}$ in a block matrix form and we only 
need to diagonalize a $(6L_{s}^{3}\times 6L_{s}^{3})$ matrix;
the chiral limit is straightforward since it only consists in 
diagonalizing $P(U;m_q=0)$.

Alternatively, one could directly diagonalize the fermionic matrix, 
expressing the determinant  as a polynomial in the mass. In this way
one is allowed to freely move in the fermionic mass, and this can be useful
for the determination of chiral observables, but simulations have to be 
repeated for each value of the chemical potential. Also, this method 
tends to be more computer demanding, since it requires the diagonalization
of  larger matrices.

The partition function (1) is real, and positive definite; 
in particular, the average fermionic determinant
at fixed energy is positive definite. This is no longer the case even in large
but finite statistics: while the average determinant can be made real, its 
sign is not definite \cite{GOCK1}. 
Therefore, and according to the previous discussion, we have chosen to compute 
$\cal Z$ by taking   the absolute value of the fermionic determinant on a 
configuration by configuration basis.
With the available statistics also other definitions we used to obtain a
positive definite partition function (like the modulus of the averaged
coefficients or the modulus of the real part of the coefficients, instead
of the average of the modulus) gives results that can not be distinguished 
among them and with the previous definition.

We have performed simulations in $4^4$, $6^3\times 4$ and $8^3\times 4$ 
lattices,  at $\beta=0$
with various values of fermion masses, starting  from $m_q=0$. 
The analysis we report in this letter  only
concerns the behaviour of the baryonic number density $N(\mu)$ and that of
its derivative with respect to $\mu$, which in the thermodynamical limit
is proportional to the  
radial 
density of the zeros of the partition function. A more comprehensive analysis, 
reporting  in particular the behaviour of chiral parameters is under way and
will be published elsewhere.

The results for $\beta=0$, $m_q=0.1$, in
$4^4$ and $6^3\times 4$ lattices are reported in Figs. 1, 2. The values of 
the parameters have been chosen to allow a direct comparison with 
\cite{BARKOG,KARSCH}. 
Several comments are in order
\begin{enumerate}
\item In the smaller volume, neither $N(\mu)$ nor the radial density indicate
a critical behaviour, apart from the (unphysical) behaviour near the
onset threshold $\mu_o=0.31$.
The small $\mu$ behaviour is unchanged in the larger volume, but a structure
develops at $\mu_c=0.69$ as well as $\mu_s\simeq 0.96$ ,
which  indicates a  phase transition in the same position as that 
found by Karsch et al. \cite{KARSCH} as well as a (first order) 
saturation transition, for which we lack of physical explanations. 
This contraddicts the hypotesys 
\cite{GOCK2}
that taking the modulus of the determinant washes out the transition.

\item Our results are in very 
good agreement with others obtained with different 
methods. In particular, up to $\mu_c$ we are in striking agreement with 
Barbour et al. \cite{BARKOG}, reproducing both the ``unphysical'' behaviour 
at small $\mu$ and the transition at $\mu_c$.
The value of $\mu_c$ agrees with that obtained in \cite{BARKOG},
but our discontinuity is  steeper, corresponding to a higher
peak in the radial distribution of zeros (see Fig. 2).

\item The value $\mu_o$ at which $N(\mu)$ departs from zero coincides with 
that found in \cite{BARKOG}. 
Contrary to $\mu_c$, however, the peak at  $\mu_o$  of the 
radial density of zeros does not drastically increase with the volume, as 
expected for a critical behaviour. We tend to exclude a first order phase 
transition at $\mu_o$.

\item The saturation at $N(\mu)=1$ is not reached smoothly in the larger 
volumes, indicating a transition at $\mu_s\simeq 0.96 $. 
Also the radial density
of zeros reported in \cite{BARKOG} shows a peak consistent with the 
existence of this transition.
\end{enumerate}

The behaviour sketched above permains for values of fermion mass as low as 
$0.02$. At lower masses the behaviour of the observables becomes smooth
in $4^4$ and $6^3\times 4$ lattices.
In Fig. 3 we report preliminary results of a simulation at $m_q=0$, 
$\beta=0$ in a $8^3\times 4$ for the radial density of the zeros 
superimposed with the same data for the smaller lattices. 
In this lattice 
a clear signal of transition develops at $\mu_c=0.65$ as well as near 
saturation. This behaviour strongly suggests that the claim \cite{GOCK2}, 
that taking the 
modulus of the fermion determinant destroys the transition, is unjustified and
what is observed is indeed a volume
effect, as function of the fermionic mass, 
disappearing consistently in larger lattices. 
It is interesting to notice that (although more statistics is
needed) the value of $\mu_c$ here obtained agrees with the chiral limit 
extrapolation of the value obtained in \cite{KARSCH}.

In fig.4 we report the dependence of $\mu_o$, $\mu_c$ and $\mu_s$ on
the quark mass. One could expect that the
partition function computed with the modulus of the determinant 
contains in the spectrum meson states with non zero baryonic charge,
as in a theory with adjoint fermions in addition to usual ones.
In this case one could expect a link between the features in
$N(\mu)$ and the mass of the lightest of such states, degenerated with the
ordinary pion \cite{DAVIES}. In the figure we also plot the strong 
coupling/mean field value of half the mass of the pion \cite{KLUBER}.
Looking at the figure we can see that the onset $\mu_o$ follows rather closely
the pion mass, at least for $m_q<0.3$; for larger values of $m_q$ a
clear displacement can be seen. Moreover the exact values of $\mu_o$ are better
described, in the former region, by a simpler relation, i.e. the square root 
of the quark mass. 
If colourless diquark states, with non zero baryonic charge, 
lighter than the nucleon exist
we would expect a saturation transition as soon as the chemical potential
is of the order of magnitude of (half) the mass of the particle.
Our data for $N(\mu)$ seem to discard this scenario; the 
(almost) linear behaviour
between onset and the transition is not modified in the largest lattice.

We have also considered the quenched case where 
no signals of transition have been found \cite{KOG}. 
In this case we compute the
derivative of $N(\mu)$, which is given by the distribution of the
eigenvalues of the propagator matrix, and
shows no sharp peaks for any  $\mu$ ($6^3\times 4$). The distribution
is reported
in Fig. 5, superimposed with that of the full theory; 
the inclusion of 
dynamical fermions is therefore crucial for the revealing of 
the transition.

In the full theory at large masses we have found  results inconsistent 
with expectations \cite{KARSCH}. 
The signal of the phase transition vanishes for
$m_q > 0.7$ and  $N(\mu)$ goes smoothly from zero to one except for 
two structures
at the onset and saturation point; this is consistent
with what reported
in \cite{BARKOG}.

To understand the results for large masses, we performed
simulations in the smaller lattice ($4^4$)
diagonalizing directly the $\Delta(m_q=0)$
matrix for several values of $\mu$. 
This way $\cal Z$ can be computed  (essentially) continuously in 
the mass. The results in the small mass range are consistent with the ones
obtained from the eigenvalues of the propagator matrix, {\it i.e.} 
no transition is found in the same lattice. 
At larger values of mass, instead, a clear transition signal appears 
and, as expected on phenomenological
grounds, it shows a saturation behaviour at a value of the chemical potential
 consistent with the 
large mass extrapolation of the results of \cite{KARSCH}. 
This is shown in Fig. 6 for $m_q=1.5$;
in the same figure we also report 
preliminary results for the chiral
condensate.

In conlusion, the situation is still far from clear.
We have shown that  
using the modulus of the determinant we can reproduce the results obtained
with other algorithms and this is the main motivation of this letter.
However, the behaviour from the onset to the  
transition at $\mu_c$ is difficult to understand in physical terms, as is
the significance of the onset;
we expect to get more informations from the analysis of chiral
observables that we will present in a future publication. 
For large masses the technique based on the propagator matrix seems
to  suffer
from numerical instabilities which may wash out the transition signal.
On the other hand, the simulations performed diagonalizing directly
the fermionic matrix appear to give results in agreement with physical
intuition.

\vskip 0.2 truecm
This work has been partly supported through a CICYT (Spain) - INFN (Italy)
collaboration. The authors thank Maria Paola Lombardo for useful 
discussions.

\newpage
\vskip 4cm
\begin{figure}[htb]
\psrotatefirst
\psfig{figure=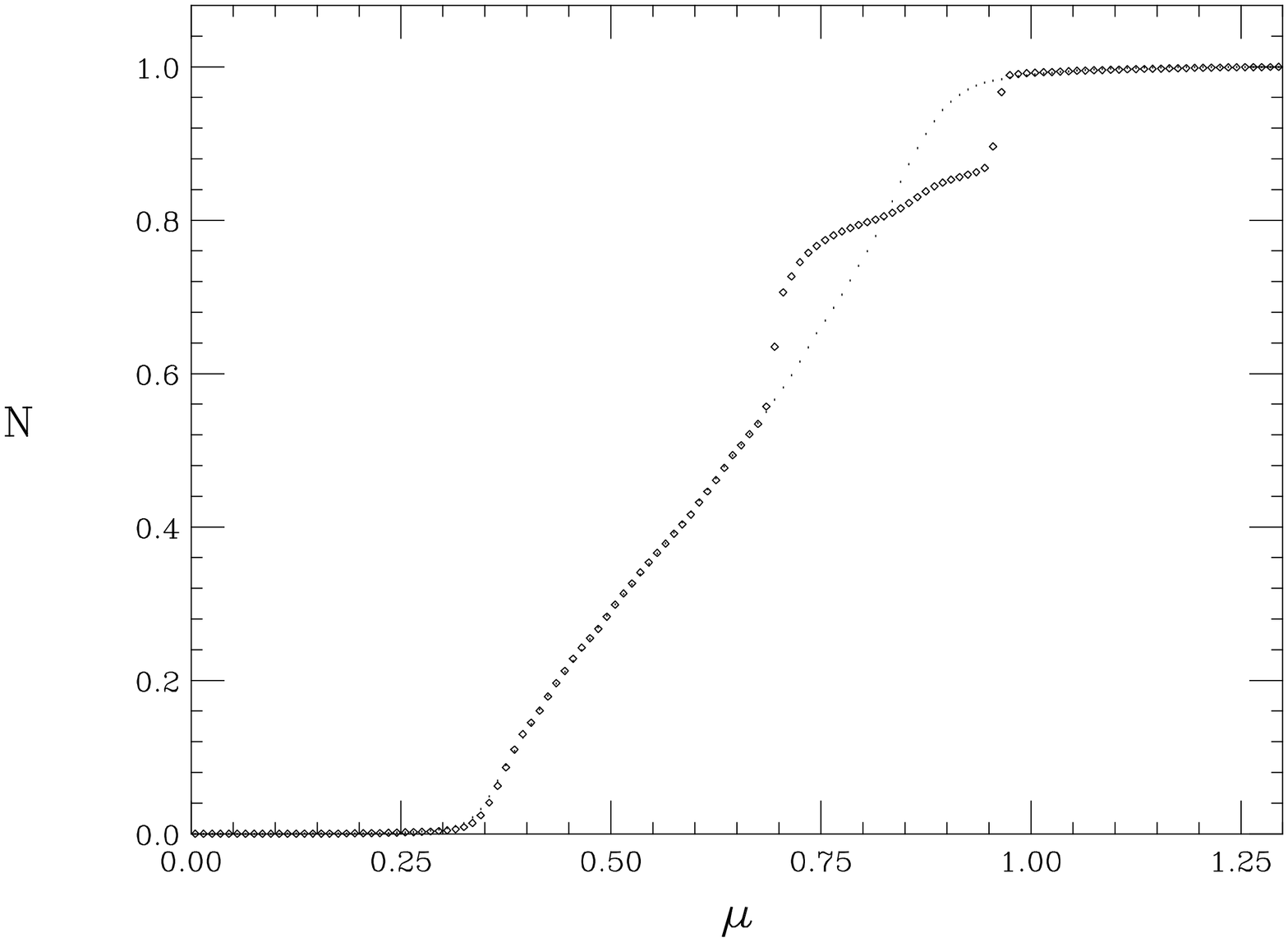,angle=90,width=16cm,height=16cm}
\caption{Number density vs. chemical potential in a $4^4$ (dots)
and $6^3 \times 4$ (diamonds) lattice at $m_{q}=0.1$}
\end{figure}
\newpage
\vskip 4cm
\begin{figure}[htb]
\psrotatefirst
\psfig{figure=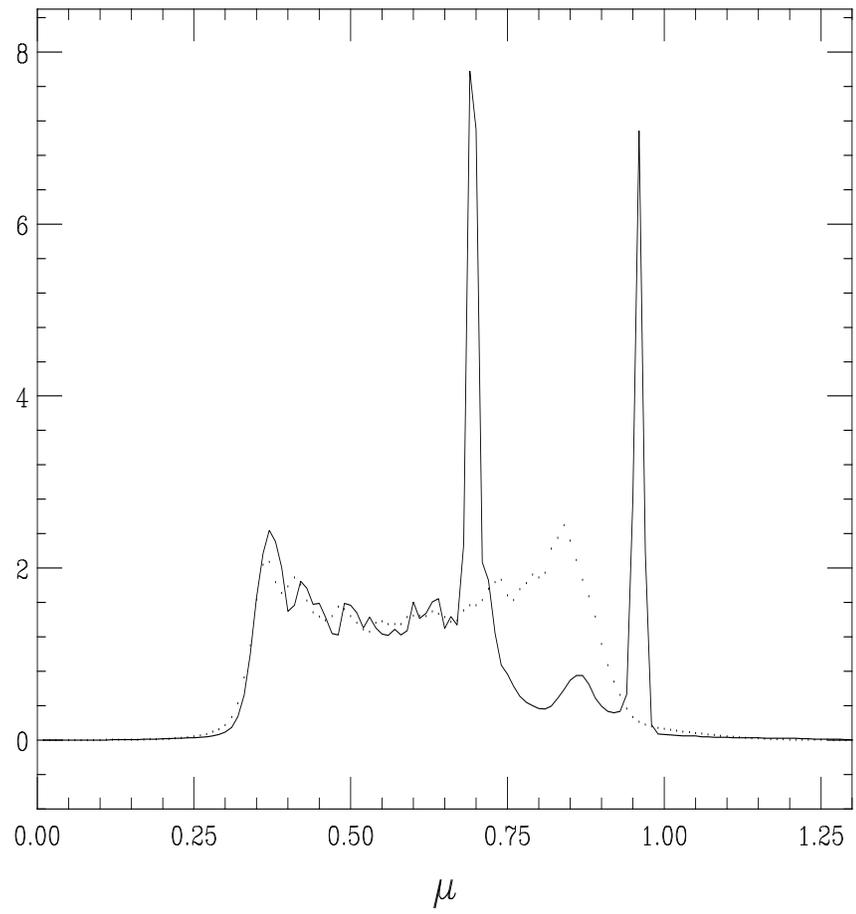,angle=90,width=16cm,height=16cm}
\caption{Radial density of zeros in a $4^4$ (dots) and $6^3 \times 4$ 
(continuous line) lattices
at $m_{q}=0.1$}
\end{figure}
\newpage
\vskip 4cm
\begin{figure}[htb]
\psrotatefirst
\psfig{figure=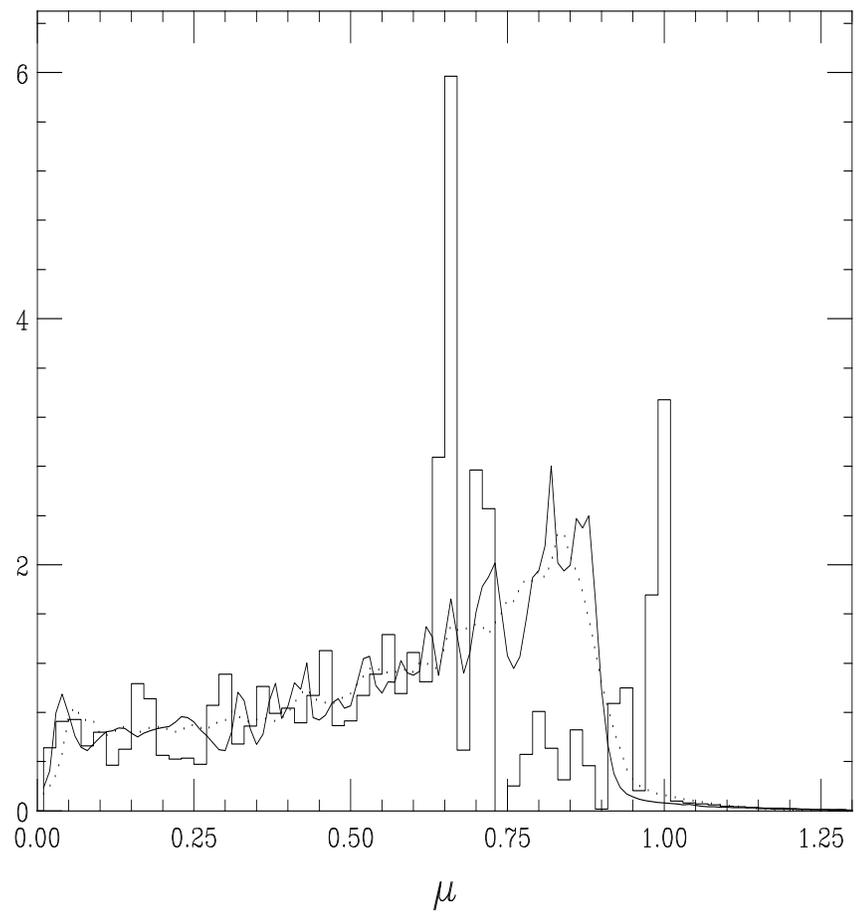,angle=90,width=16cm,height=16cm}
\caption{Radial density of zeros in a $4^4$ (dots), $6^3 \times 4$ (continuous
line) and $8^3\times 4$ (histogram) lattices at $m_q=0.0$}
\end{figure}
\newpage
\vskip 4cm
\begin{figure}[htb]
\psrotatefirst
\psfig{figure=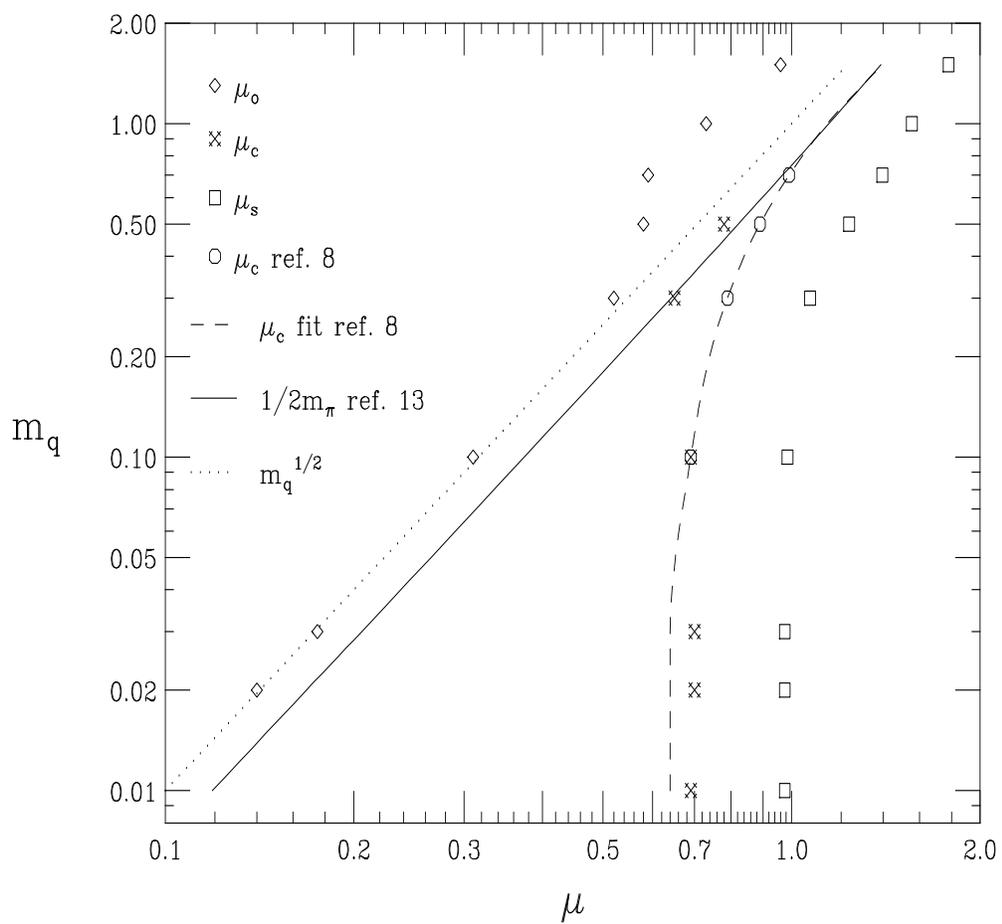,angle=90,width=16cm,height=16cm}
\caption{$\mu_o,\mu_c$ and $\mu_s$ in the $(\mu,m_q)$ plane in a 
$6^3 \times 4$ lattice}    
\end{figure}
\newpage
\vskip 4cm
\begin{figure}[htb]
\psrotatefirst
\psfig{figure=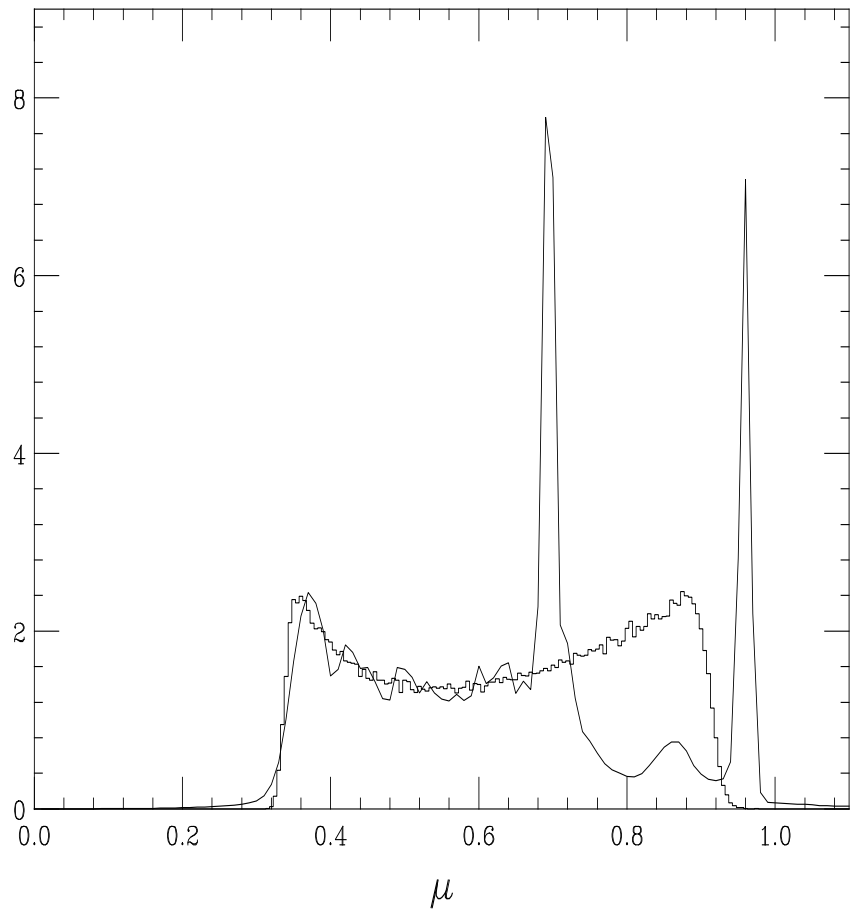,angle=90,width=16cm,height=16cm}
\caption{Radial density of zeros quenched (histogram) and unquenched 
(continuous line) in a $6^3\times 4$ lattice at $m_q=0.1$.}
\end{figure}
\newpage
\vskip 4cm
\begin{figure}[htb]
\psrotatefirst
\psfig{figure=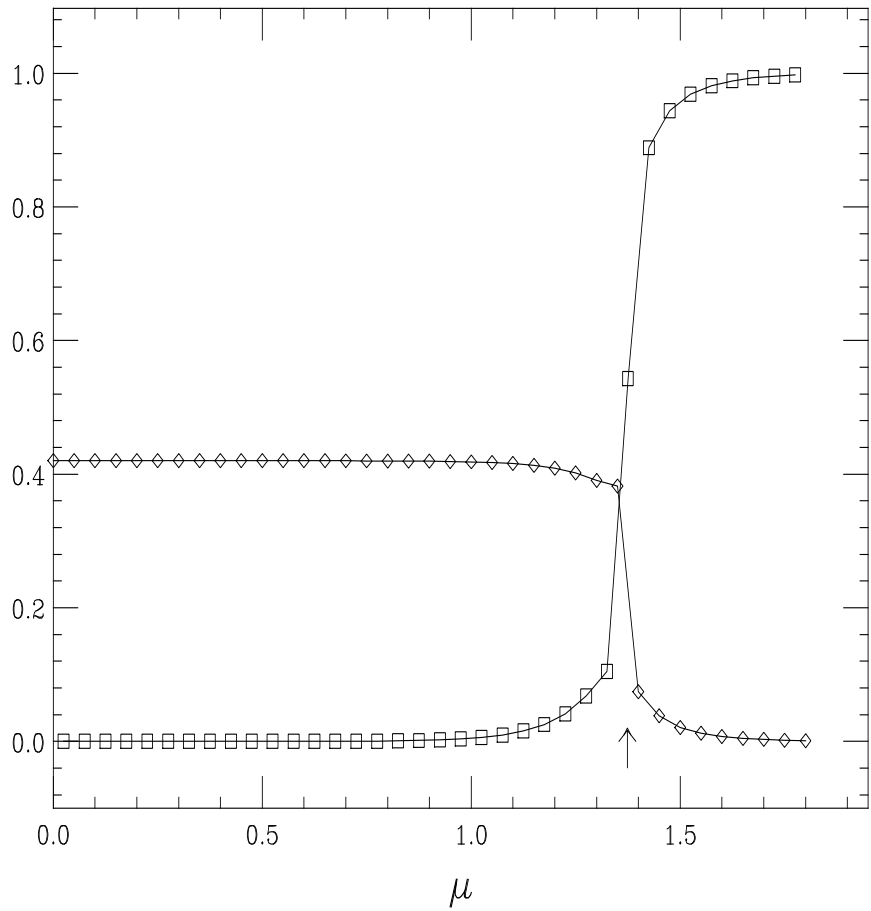,angle=90,width=16cm,height=16cm}
\caption{Number density (squares) and chiral condensate (diamonds) vs.
chemical potential in a $4^4$ lattice at $m_q=1.5$. The arrow indicates 
the transition point obtained from an extrapolation of the data
in \cite{KARSCH}.}
\end{figure}

\end{document}